\documentclass[11pt]{article}
\usepackage{amsmath,amsthm,amsfonts}
\usepackage[usenames]{color}
\theoremstyle{definition}

\theoremstyle{definition}

\theoremstyle{definition}
 
\def\G{\Gamma}

\begin{document}

\title{\LARGE\bf Gauge Dependence in the Nonlinearly Realized\\ 
Massive $SU(2)$ Gauge Theory}
\date{}
\author{\Large D.~Bettinelli, R.~Ferrari, A.~Quadri\\ \\
Department of Physics, University of Milan\\
and INFN, Sez. di Milano\\
via Celoria 16, I-20133 Milano, Italy\\ \\
E-mails: daniele.bettinelli@mi.infn.it, ruggero.ferrari@mi.infn.it, andrea.quadri@mi.infn.it}
\maketitle
\thispagestyle{empty}
\begin{abstract}
The implementation of the 't Hooft $\alpha$-gauge in the
symmetrically subtracted  massive gauge theory based on the nonlinearly
realized $SU(2)$ gauge group is discussed. 
The gauge independence of the self-mass of the gauge
bosons is proven by cohomological techniques.
\par\smallskip
Preprint number: IFUM-908-FT
\end{abstract}

\section{Introduction}

A consistent subtraction scheme for massive non-Abelian
gauge theories based on a nonlinearly realized gauge group has been
recently proposed in \cite{Bettinelli:2007tq}. 
The symmetric subtraction algorithm was already successfully applied 
to the 
four-dimensional nonlinear sigma model  in the flat
connection formalism in Refs.\cite{Ferrari:2005ii}-\cite{Bettinelli:2007zn}.

The Feynman rules of the nonlinearly realized massive gauge theory
entail that already at one loop level there is an infinite number 
of divergent amplitudes involving
the pseudo-Goldstone fields $\phi_a$ \cite{Bettinelli:2007tq}. 
The latter amplitudes are uniquely fixed 
by implementing a defining local functional equation \cite{Bettinelli:2007tq,
Ferrari:2005ii}
which encodes the invariance of the path-integral Haar measure 
under local $SU(2)_L$ transformations\footnote{the subscript $L$ stands 
for the left action on the group element.}
\begin{eqnarray}
&& \Omega' = U_L \Omega \, , \cr
&& A'_\mu = U_L A_\mu U_L^\dagger + i U_L
\partial_\mu U_L^\dagger \, .
\label{eq.0}
\end{eqnarray}
$\Omega$ is the element of the nonlinearly represented $SU(2)_L$ gauge
group
\begin{eqnarray}
\Omega = \frac{1}{v_D} (\phi_0 + i \phi_a \tau_a) \, , ~~~~~  \phi_0 = \sqrt{v_D^2 - \phi_a^2} \, .
\label{eq.0.1}
\end{eqnarray}
$v_D$ is a $D$-dimensional mass scale and $\tau_a$ are
the Pauli matrices. $A_\mu = A_{a\mu} \frac{\tau_a}{2}$ is the $SU(2)_L$
gauge connection.
It is also convenient to introduce the $SU(2)_L$ flat connection
\begin{eqnarray}
F_\mu = F_{a\mu} \frac{\tau_a}{2} = i \Omega \partial_\mu \Omega^\dagger \, ,
\label{eq.0.2}
\end{eqnarray}
with the following $SU(2)_L$ transformation induced by the 
transformation of $\Omega$
\begin{eqnarray}
F'_\mu =  U_L F_\mu U_L^\dagger + i U_L
\partial_\mu U_L^\dagger \, .
\label{eq.0.3}
\end{eqnarray}
The amplitudes not involving the pseudo-Goldstone fields
are named ancestor amplitudes since they are at the top
of the hierarchy induced by the local functional equation.
At every loop order there is only a finite number of 
divergent ancestor amplitudes (weak power-counting theorem
\cite{Bettinelli:2007tq},\cite{Ferrari:2005va}).
The requirement
of physical unitarity is satisfied since the
Slavnov-Taylor (ST) identity holds 
\cite{Bettinelli:2007tq},\cite{Ferrari:2004pd}. 
The ghost equation and
the Landau gauge equation are also preserved by the
symmetric subtraction \cite{Bettinelli:2007tq}.
These symmetries, supplemented by global $SU(2)_R$ invariance 
and the weak power-counting, uniquely fix the tree-level vertex
functional of the nonlinearly realized theory \cite{Bettinelli:2007tq}.

In \cite{Bettinelli:2007tq} the Landau gauge was used
for the sake of simplicity and conciseness. 
The aim of this note is to implement the 't Hooft $\alpha$-gauge
in a way compatible with all the symmetries required for
the definition of the model (local functional equation,
ST identity, ghost equation, $B$-equation for a general $\alpha$-gauge
\cite{Piguet:1995er}) and the weak power-counting.

The validity of the local functional equation requires 
that the gauge-fixing functional transforms in the adjoint
representation of $SU(2)_L$.
In the Landau gauge this was achieved by introducing an external
vector source $V_{a\mu}$ and by making use of the gauge-fixing
functional
\begin{eqnarray}
\int d^Dx \, B_a (D^\mu[V] (A-V)_\mu)_a 
\label{eq.1}
\end{eqnarray}
where $(D_\mu[V])_{ac} = \partial_\mu \delta_{ac} + \epsilon_{abc} V_{b\mu}$
is the covariant derivative w.r.t. the vector source $V_{a\mu}$.
$B_a$ is the Nakanishi-Lautrup field \cite{Piguet:1995er}.
It transforms in the adjoint representation of $SU(2)_L$.
The local functional equation 
is preserved by the gauge-fixing (\ref{eq.1}).

It should be stressed that the local functional equation 
associated with the $SU(2)_L$ local invariance
is not the standard background Ward identity 
\cite{Abbott:1980hw}-\cite{Ferrari:2000yp}.
The essential difference is that the local functional equation
is bilinear in the vertex functional $\G$, due to the presence
of the nonlinear constraint $\phi_0$ in eq.(\ref{eq.0.1}),
which needs to be coupled to the scalar source $K_0$ in the tree-level vertex
functional.

The 't Hooft $\alpha$-gauge is defined by the condition
of the cancellation 
(once the Nakanishi-Lautrup field is eliminated via
its equation of motion) of the mixed $A_\mu\phi$ terms 
arising in the nonlinear theory from the mass invariant
\begin{eqnarray}
\int d^Dx \, \frac{M^2}{2} ( A_{a\mu} - F_{a\mu})^2 \, .
\label{eq.2}
\end{eqnarray}
The 't Hooft ($\phi$-dependent) gauge-fixing functional must 
transform in the adjoint representation of 
$SU(2)_L$  in order to preserve the 
local functional equation.
For that purpose one needs to introduce 
an auxiliary matrix $\widehat \Omega$
\begin{eqnarray}
\widehat \Omega = \frac{1}{v_D} ( \widehat \phi_0 + i
\widehat \phi_a \tau_a) \, 
\label{eq.3}
\end{eqnarray}
with the same $SU(2)_L$ transformation as $\Omega$:
\begin{eqnarray}
\widehat \Omega' = U_L \widehat \Omega \, .
\label{eq.4}
\end{eqnarray}
The combination
\begin{eqnarray}
&& {\cal F} = {\cal F}_a \frac{\tau_a}{2} \, , ~~~~~ 
{\cal F}_a = D^\mu[V](A-V)_{a\mu} + \frac{M^2}{2\alpha} 
                Tr [ i \widehat \Omega^\dagger \tau_a \Omega + h.c ] \, 
\label{eq.5}
\end{eqnarray}
has the correct transformation properties.
Therefore one can consider the following gauge-fixing
functional
\begin{eqnarray}
\int d^Dx \, \Big [ - \frac{1}{2\alpha} B_a^2 + B_a {\cal F}_a \Big ] \, ,
\label{eq.6}
\end{eqnarray}
where $\alpha$ is the gauge parameter.
With the choice in eq.(\ref{eq.6}) the local functional equation is respected
and by integrating the Nakanishi-Lautrup field $B_a$ 
the mixed $A_\mu\phi$-terms arising from eq.(\ref{eq.2}) are canceled.
The propagators obtained by using the gauge-fixing functional
in eq.(\ref{eq.6}) have a UV behaviour compatible with the weak
power-counting.

It is important to realize that 
$\widehat \Omega$ is not an element of $SU(2)$. 
The reason is that the amplitudes involving $\widehat \phi_0$ and
$\widehat \phi_a$ must be ancestors. 
Already at one loop level
one cannot have a finite number of divergent amplitudes
involving $\widehat \phi_a$ if $\widehat \phi_0$ is 
given by the $SU(2)$ constraint $\widehat \phi_0^2 + \widehat \phi_a^2 = v_D^2$.
One must split in a linear way 
the constant component of $\widehat \phi_0$ by setting
$\widehat \phi_0 \equiv v_D + \widehat \sigma$.
Since $\widehat \sigma$ and $\widehat \phi_a$ are 
independent, by inspecting the Feynman rules one can then
check that the weak power-counting is preserved.
Moreover, since $\widehat \phi_0, \widehat \phi_a$
are independent variables, the BRST transformation can be extended to
these sources by pairing them to external source ghosts $\theta_0,
\theta_a$ as follows
\begin{eqnarray}
&& s \widehat \phi_0 = \theta_0 \, , ~~~~ 
s \theta_0 = 0 \, , ~~~ 
s \widehat \phi_a = \theta_a \, ,
 ~~~~ s \theta_a = 0 \, .
\label{eq.7}
\end{eqnarray}
Then $(\widehat \phi_0, \theta_0)$, $(\widehat \phi_a, \theta_a)$
form BRST doublets \cite{Piguet:1995er,Quadri:2002nh}
and therefore they do not contribute to the cohomology $H(s)$ 
of the BRST differential $s$. Hence they are not physical,
as expected. The same technique can be used to prove
that  the vector source $V_{a\mu}$ does not modify the physical
observables too \cite{Bettinelli:2007tq}.

The ghost-antighost part of the tree-level vertex functional is 
generated as usual by 
\begin{eqnarray}
\int d^Dx \, s \Big [ \bar c_a ( - \frac{1}{2\alpha}
B_a + {\cal F}_a ) \Big ] = 
\int d^Dx \, \Big ( - \frac{1}{2\alpha} B_a^2 
+ B_a {\cal F}_a - \bar c_a s {\cal F}_a \Big ) \, ,
\label{eq.8}
\end{eqnarray}
with the standard BRST transformation of the antighost $\bar c_a$, i.e.
$s \bar c_a = B_a \, , s B_a = 0$.

\section{Gauge dependence of the physical amplitudes}

The issue arises of whether physical amplitudes depend
on the gauge choice. This problem can be
treated according to the standard extension of the
BRST symmetry to the gauge parameter 
\cite{Piguet:1995er,Piguet:1984js}
\begin{eqnarray}
s \alpha = \zeta \, , ~~~ s \zeta = 0 \, .
\label{eq.10}
\end{eqnarray}
The extended ST identity is in fact sufficient to prove the
independence of the physical quantities from $\alpha$ also
in the nonlinear case.
We sketch here the main lines of the proof.
Dropping inessential terms involving the background ghosts
$\theta_0,\theta_a$ the ST identity for the vertex functional $\G$
is\footnote{$\G_X$ denotes the functional derivative of $\G$ w.r.t. $X$.}
\begin{eqnarray}
{\cal S}(\G) = \int d^Dx \, \Big [ \G_{A^*_{a\mu}} \G_{A^\mu_a} 
+ \G_{\phi_a^*} \G_{\phi_a} + \G_{c^*_a} \G_{c_a} + B_a
\G_{\bar c_a} \Big ] + \zeta \frac{\partial \G}{\partial \alpha} = 0 \, .
\label{eq.11}
\end{eqnarray}
We define as usual the connected generating functional $W$
by the Legendre transformation of $\G$ w.r.t. the quantized
fields (collectively denoted by $\underline{\Phi}$)
\begin{eqnarray}
W= \G + \int d^Dx \, 
\underline{K} ~ \underline{\Phi} \, , 
\label{eq.12}
\end{eqnarray}
where $\underline{K}$  
is a short-hand notation for the sources of the quantized fields.
Eq.(\ref{eq.11}) yields ($K(\varphi)$ stands for the source
coupled to the field $\varphi$)
\begin{eqnarray}
{\cal S}(W) = \int d^Dx \, \Big [ - K(A_{a\mu}) W_{ A^*_{a\mu}}
- K(\phi_a) W_{\phi_a^*} - K(c_a) W_{c^*_a} - W_{K(B_a)} K(\bar c_a)
\Big ] 
+ \zeta \frac{\partial W}{\partial \alpha} = 0 \, .
\label{eq.13}
\end{eqnarray}
By differentiating eq.(\ref{eq.13}) w.r.t. $\zeta$ and a set of
sources $\beta_1, \dots, \beta_n$ coupled to physical BRST-invariant 
local operators ${\cal O}_1, \dots, {\cal O}_n$ 
one finds by going on-shell (all external sources
set to zero)
\begin{eqnarray}
\left . \frac{\partial}{\partial \alpha} W_{\beta_1 \dots \beta_n} 
\right |_{on-shell}= 0 \, ,
\label{eq.14}
\end{eqnarray}
i.e. the physical Green function $W_{\beta_1 \dots \beta_n}$
is on-shell gauge-independent.

The Nielsen identities \cite{Nielsen:1975fs,Gambino:1999ai}
can also be obtained from the extended ST identity (\ref{eq.11}). 
We discuss here in detail the Nielsen identity for the
two point 1-PI function of the gauge bosons.

By differentiating eq.(\ref{eq.11}) w.r.t. $A_{a_1 \mu_1}, A_{a_2 \mu_2}$
and $\zeta$ and by setting all the fields and external sources to zero
one gets
\begin{eqnarray}
&& \G_{\zeta A^*_{a\mu} A_{a_1 \mu_1}} \G_{A_{a_2 \mu_2} A_{a\mu}} +
\G_{\zeta A^*_{a\mu} A_{a_2 \mu_2}} \G_{A_{a_1 \mu_1} A_{a\mu}} + \nonumber \\
&& \G_{\zeta \phi^*_a A_{a_1 \mu_1}} \G_{A_{a_2 \mu_2} \phi_a} +
\G_{\zeta \phi^*_a A_{a_2 \mu_2}} \G_{A_{a_1 \mu_1} \phi_a} +
\partial_\alpha \G_{A_{a_1 \mu_1} A_{a_2 \mu_2}} = 0 \, .
\label{eq.15}
\end{eqnarray}
We decompose $\G_{AA}$ and $\G_{\zeta A^* A}$ into their
transverse and longitudinal components as follows
\begin{eqnarray}
&& \G_{A_{a\mu}(-p) A_{b\nu}(p)} = \delta_{ab} \Big ( \Sigma^{AA}_T(p^2) T_{\mu\nu}
+ \Sigma^{AA}_L(p^2) L_{\mu\nu} \Big ) \, , \nonumber \\
&&
\G_{\zeta A^*_{a\mu}(-p) A_{b\nu}(p)} = \delta_{ab}
\Big ( \Sigma_T^{\zeta A^* A}(p^2) T_{\mu\nu} + \Sigma^{\zeta A^* A}_L(p^2) L_{\mu\nu} \Big ) \, , \nonumber \\
&& T_{\mu\nu} = g_{\mu\nu} - \frac{p_\mu p_\nu}{p^2} \, , 
~~~~ L_{\mu\nu} = \frac{p_\mu p_\nu}{p^2} \, .
\label{eq.16}
\end{eqnarray}
By taking the transverse part of eq.(\ref{eq.15}) one finds
(notice that the terms proportional to
$\G_{A_\mu \phi}$ only contribute to the longitudinal part and
thus drop out)
\begin{eqnarray}
\partial_\alpha \Sigma^{AA}_T(p^2) = - 2 \Sigma_T^{\zeta A^* A}(p^2) \Sigma_T^{AA}(p^2) 
\, .
\label{eq.17}
\end{eqnarray}
The self-mass $\overline{M}^2$ is defined as the zero of 
$\Sigma^{AA}_T(p^2)$:
\begin{eqnarray}
\Sigma^{AA}_T(\overline{M}^2)=0 \,  
\label{eq.18}
\end{eqnarray}
if no tadpoles are present as in the  case under consideration. 
In passing it is worth noticing that the presence of tadpoles
requires that the self-mass is defined as the pole of the
transverse part of the
connected two-point function (as it happens in the linearly realized
theory).

By eq.(\ref{eq.17}) one finds 
\begin{eqnarray}
\partial_\alpha \Sigma^{AA}_T(\overline{M}^2) = 0 \, .
\label{eq.19}
\end{eqnarray}
Moreover invertibility of $\frac{\partial \Sigma^{AA}_T}{\partial p^2}$
is guaranteed in the loop expansion since
$$\frac{\partial \Sigma^{AA}_T}{\partial p^2} = -1 + O(\hbar) \, .$$
The above equation together with eq.(\ref{eq.19})
implies 
\begin{eqnarray}
\frac{\partial \overline{M}^2}{\partial \alpha} = 0 \, ,
\label{eq.20}
\end{eqnarray}
i.e. the self-mass is gauge-independent. This behaviour
is a typical property of the nonlinear theory. In the linear case
the zero of the two-point 1-PI function
 is in general gauge-dependent. Gauge independence
can only be recovered by taking into account the
Higgs tadpole contributions. For a detailed
comparison of the two-point 1-PI function in the
linear and the nonlinear case see Ref.~\cite{Bettinelli:2007cy}.

\section{Conclusions}

The formulation of the nonlinearly realized $ SU(2)$ massive gauge
theory in the 't Hooft gauge has been achieved in a way
consistent with all the symmetries of the model and the
weak power-counting. This requires the introduction of auxiliary
external sources $\widehat \sigma, \widehat \phi_a$. 
We have shown that this procedure
does not alter the physical content of the model. 
Gauge independence of physical observables has been established
by using cohomological methods. 
The self-mass, which can be
computed in the nonlinearly realized theory 
as the zero of the transverse part of the 
1-PI two-point function, has been proven to be gauge-independent.

\section*{Acknowledgement}
This work was partially supported by INFN.


\begin{thebibliography}{99}
\itemsep-4pt

\bibitem{Bettinelli:2007tq}
  \newblock{D.~Bettinelli, R.~Ferrari and A.~Quadri,}
  \newblock{\it A massive Yang-Mills Theory based on the nonlinearly 
   realized gauge group,}
  \newblock{arXiv:0705.2339 [hep-th].}

\bibitem{Ferrari:2005ii}
  \newblock{R.~Ferrari,}
  \newblock{\it Endowing the nonlinear sigma model with a 
             flat connection structure: A  way to renormalization,}
  \newblock{JHEP {\bf 0508} (2005) 048 [arXiv:hep-th/0504023].}

\bibitem{Ferrari:2005va}
  \newblock{R.~Ferrari and A.~Quadri,}
  \newblock{\it A weak power-counting theorem for the renormalization of the non-linear sigma model in four dimensions,}
  \newblock{Int.\ J.\ Theor.\ Phys.\  {\bf 45} (2006) 2497
  [arXiv:hep-th/0506220].}

\bibitem{Ferrari:2005fc}
  \newblock{R.~Ferrari and A.~Quadri,}
  \newblock{\it Renormalization of the non-linear sigma model in four dimensions: A two-loop example,}
  \newblock{JHEP {\bf 0601} (2006) 003
  [arXiv:hep-th/0511032].}

\bibitem{Bettinelli:2007zn}
  \newblock{D.~Bettinelli, R.~Ferrari and A.~Quadri,}
  \newblock{\it A comment on the renormalization of the nonlinear sigma model,}
  \newblock{arXiv:hep-th/0701197, to appear in Int.\ J.\ Mod.\ Phys.\ A}.

\bibitem{Ferrari:2004pd}
  R.~Ferrari and A.~Quadri,
  \newblock{\it Physical unitarity for massive non-Abelian gauge theories in the Landau gauge: Stueckelberg and Higgs,} 
  \newblock{JHEP {\bf 0411} (2004) 019 [arXiv:hep-th/0408168].}

\bibitem{Piguet:1995er}
  \newblock{O.~Piguet and S.~P.~Sorella,}
  \newblock{\it 
   Algebraic renormalization: Perturbative renormalization, symmetries and
   anomalies,}
  Lect.\ Notes Phys.\  {\bf M28}, Springer 1995.

\bibitem{Abbott:1980hw}
  \newblock{L.~F.~Abbott,}
  \newblock{\it The background field method beyond one loop,}
  \newblock{Nucl.\ Phys.\  B {\bf 185} (1981) 189.}

\bibitem{Abbott:1983zw}
  \newblock{L.~F.~Abbott, M.~T.~Grisaru and R.~K.~Schaefer,}
  \newblock{\it The background field method and the S matrix,}
  \newblock{Nucl.\ Phys.\  B {\bf 229} (1983) 372.}

\bibitem{Becchi:1999ir}
  \newblock{C.~Becchi and R.~Collina,}
  \newblock{\it Further comments on the background field method and gauge invariant
  effective actions,}
  \newblock{Nucl.\ Phys.\  B {\bf 562}, 412 (1999) [arXiv:hep-th/9907092].}

\bibitem{Ferrari:2000yp}
  \newblock{R.~Ferrari, M.~Picariello and A.~Quadri,}
  \newblock{\it Algebraic aspects of the background field method,}
  \newblock{Annals Phys.\  {\bf 294}, 165 (2001)
  [arXiv:hep-th/0012090].}

\bibitem{Quadri:2002nh}
  \newblock{A.~Quadri,}
  \newblock{\it Algebraic properties of BRST coupled doublets,}
  \newblock{JHEP {\bf 0205} (2002) 051 [arXiv:hep-th/0201122].}

\bibitem{Piguet:1984js}
  \newblock{O.~Piguet and K.~Sibold,}
  \newblock{\it Gauge independence in ordinary Yang-Mills theories,}
  \newblock{Nucl.\ Phys.\  B {\bf 253} (1985) 517.}

\bibitem{Nielsen:1975fs}
  \newblock{N.~K.~Nielsen,}
  \newblock{\it On the gauge dependence of spontaneous symmetry breaking in gauge theories,}
  \newblock{Nucl.\ Phys.\  B {\bf 101} (1975) 173.}

\bibitem{Gambino:1999ai}
  \newblock{P.~Gambino and P.~A.~Grassi,}
  \newblock{\it The Nielsen identities of the SM and the definition of mass,}
  \newblock{Phys.\ Rev.\  D {\bf 62}, 076002 (2000)
  [arXiv:hep-ph/9907254].}

\bibitem{Bettinelli:2007cy}
  \newblock{D.~Bettinelli, R.~Ferrari and A.~Quadri,}
  \newblock{\it One-loop self-energy and counterterms in a massive Yang-Mills Theory based on the nonlinearly realized gauge group,}
  \newblock{arXiv:0709.0644 [hep-th].}

\end{thebibliography}
\end{document}